\documentclass{jnmp}

\usepackage{amsmath}

\JNMPnumberwithin{equation}{section}

\theoremstyle{definition}

\begin{document}

%
\renewcommand{\evenhead}{P~G~L Leach, M~C~Nucci and S~Cotsakis}
\renewcommand{\oddhead}{Symmetry, Singularities and Integrability in Complex
Dynamics}

%
\thispagestyle{empty}

\FirstPageHead{8}{4}{2001}{\pageref{leach-firstpage}--\pageref{leach-lastpage}}{Article}

\copyrightnote{2001}{P~G~L Leach, M~C~Nucci and S~Cotsakis}

\Name{Symmetry, Singularities and Integrability\\
in Complex Dynamics V: Complete Symmetry Groups of Certain Relativistic
Spherically Symmetric Systems}
\label{leach-firstpage}

\Author{P~G~L LEACH~${}^\dag{}^*$, M~C~NUCCI~$^\ddag$ and S~COTSAKIS~$^\dag$}

\Address{$^\dag$~GEODYSYC, Department of Mathematics, University of the Aegean,\\
~~Karlovassi 83 200, Greece\\[10pt]
$^*$~Permanent address: School of Mathematical and Statistical Sciences,\\
~~University of Natal, Durban 4041, Republic of South Africa\\[10pt]
$^\ddag$~Dipartimento di Matematica e Informatica, Universit\`a di Perugia, 06123 Perugia, Italy}

\Date{Received August 3, 2000; First Revision December 7, 2000;
 Second Revision May 4, 2001;
Accepted May 5, 2001}

\begin{abstract}
\noindent
We show that the concept of complete symmetry group introduced by
Krause ({\it J.~Math. Phys.} {\bf 35} (1994), 5734--5748) in the
context of the Newtonian Kepler problem has wider applicability,
extending to the relativistic context of the Einstein equations
describing spherically symmetric bodies with certain conformal
Killing symmetries. We also provide a simple demonstration of the
nonuniqueness of the complete symmetry group.
\end{abstract}

\section{Introduction}

In 1994 Krause \cite{Krause} introduced the concept of a complete
symmetry group in classical mechanics.  This concept was somewhat
stronger than that of the standard symmetry group generally used
at the time, which was the group represented by the Lie point
symmetries of the system of differential equations describing the
dynamical system.  A complete symmetry group realisation in
mechanics was required to be endowed with the properties that (i)
the group act freely and transitively on the manifold of all
allowed motions of the system and (ii) the given equations of
motion be the only ordinary differential equations that remain
invariant under the specified action of the group.  The vehicle
which Krause used to demonstrate the concept was the classical
Kepler problem described in reduced coordinates by the second
order ordinary differential equation
 \begin{equation}
 \ddot{\pbf r} +\frac{\mu{\pbf r}}{r^3} = 0, \label{1.1}
 \end{equation}
where $\mu $ is a constant, $\pbf{r} = (x_1,x_2,x_3) $ and $r
=|\pbf{r}| $ in a standard notation.  The Lie point symmetries of
the system are given by
 \begin{align}
& G_1 = x_2\partial_{x_3} -x_3\partial_{x_2},       &&
G_2 = x_3\partial_{x_1} -x_1\partial_{x_3},    &
G_3 = x_1\partial_{x_2} -x_2\partial_{x_1}, \nonumber\\
& G_4 =\partial_t, &&  G_5 = 3t\partial_t + 2r\partial_r,  &  \label{1.2}
 \end{align}
where $G_1 $, $G_2 $ and $G_3 $ are the generators of rotation,
$G_4 $ is the generator of time translations and $G_5 $
represents rescaling which maps orbits of a given eccentricity
into orbits of the same eccentricity and is the basis of Kepler's
Third Law because of the happy happenstance that for the Kepler
problem (and also the simple harmonic oscillator) the
eccentricity does not enter into the relationship between the
period of the orbit and the length of the semimajor axis
as it does for other periodic orbits~\cite{Gorringe}.  The
algebra of the point symmetries is $so (3)\oplus A_2 $.  In
addition to the standard Lie point symmetries Krause determined
the three nonlocal symmetries
 \begin{gather}
 X_1 = \left(2\int x_1\mbox{d} t\right)\partial_t +x_1r\partial_r, \qquad
 X_2 = \left(2\int x_2\mbox{d} t\right)\partial_t +x_2r\partial_r , \nonumber\\
 X_3 = \left(2\int x_3\mbox{d} t\right)\partial_t +x_3r\partial_r \label{1.3}
 \end{gather}
and with these eight symmetries was able to specify the precise
form of (\ref{1.1}) up to the value of the constant $\mu $.  That
the constant $\mu $ could not be specified is not surprising
since it can be removed from the equation~(\ref{1.1}) by a
rescaling of the vector $\pbf{r}$.

The work of Krause was and is very interesting. Apart from any
other features which are peculiar to the Kepler problem, there
was presented the new approach to the calculation of nonlocal
symmetries which are not easy to calculate in an algorithmic way
\cite{Govinder}. It is unfortunate that in his paper Krause made
the comment that the complete symmetry group could not be
obtained for the Kepler problem by the standard analysis used for
Lie point symmetries.  Rising to the implied challenge
Nucci~\cite{Nucci} proceeded to devise a scheme of reduction of order
of the system, in essence a removal of time as the independent
variable so that the coefficient function of $\partial_t $ appeared
only as the derivative and hence a local function for the type of nonlocal
symmetry used by Kause, which made the
determination of the nonlocal symmetries above a process of the
determination of the Lie point symmetries of the reduced system.
More recently Nucci and Leach~\cite{Nucci2} have shown that the
number of nonlocal symmetries of this class for the Kepler
problem was somewhat greater than Krause had reported.
Furthermore, in contrast to his statement that one does not
obtain a Lie algebra when the nonlocal symmetries are considered,
they presented the algebra.

Krause emphasised that his treatment applied to mechanical systems
based on the Newtonian laws, but there can be no doubt that the
concept of a complete symmetry group is applicable to all systems
of ordinary differential equations which possess a suitable number
of symmetries to define precisely the system of equations.  Krause
further remarks that it is not known whether the specific
realisation of a complete symmetry group exists for any given
Newtonian system~--- more generally system of ordinary differential
equations~--- and that the question was open.  There is a specific
realisation for one-dimensional linear second order equations, $sl
(3,R) $, but even for nonlinear one-dimensional systems the
problem was unresolved.  This immediately raises the question of
what the situation would be in the case of dynamical systems known
to display the characteristics of nonintegrability.  Considering
that the existence of symmetry is regarded as one of the criteria
relating to integrability, the concept of the complete symmetry
group of a nonintegrable dynamical system would appear to present
something of a paradox.

In this paper, which is the fifth of a series
\cite{Flessas,II,III,IV} devoted to investigating the connections
between the three main topics in dynamics, {\it videlicet} symmetry,
singularities and integrability, topics which superficially are
unrelated in their mathematics and yet are intimately intertwined,
we demonstrate that it is possible to have a complete symmetry
group of an ordinary differential equation which does not have
sufficient Lie point symmetries for integrability and is not
integrable in the sense of Painlev\'{e}.  In the process we
demonstrate the resolution of the apparent paradox.

In the normal course of events one associates the existence of
symmetry in a differential equation (or system of differential equations)
with the integrability of the equation.  However, the normal course
of events also restricts the consideration of a symmetry to the class
 of point, contact or generalised symmetries.  In the very introduction
 of the concept of a complete symmetry group Krause was obliged to resort
 to nonlocal symmetries for the complete specification of the system of
 differential equations for the Kepler problem which is well-known to be
  an integrable system.  The need to use nonlocal symmetries for the
   complete symmetry group is not automatic.
   For example the equation $y''= 0 $ is specified completely
   by three point symmetries.

   In this paper we are not concerned
    with the existence of symmetries which will enable the equation
     to be integrated.  We are concerned with the existence of symmetries
     which will enable the equation to be specified completely, perhaps
     up to an arbitrary parameter as was the case of Krause's complete
      symmetry group for the Kepler problem.  The role played by the
      elements of the complete symmetry group is to fix the form of
      the dynamical equations.  Our aim is to demonstrate that this
      be the case whether the system be integrable or not integrable.
      As we shall see, the very nature of the symmetries obtained is
      not likely to cause one to imput integrability of the model equation.
       It would appear that symmetries can play two roles, one to specify
       the nature of the equation and the other to provide part of the
        route to integrability.  {\it A priori} one cannot expect both
         roles to the played by the one symmetry.

This paper is structured as follows.  In the next section we
review the known properties of our model equation.  In Section~3
we calculate its symmetries and demonstrate that within the
number of these symmetries there is a subset which is the
complete symmetry group of the model equation.  In Section~4 we
conclude with some observations.

\resetfootnoterule

\section{Painlev\'{e} nonintegrability}

 Our model equation is
 \begin{equation}
 y'''+y''+yy'= 0, \label{2.1}
 \end{equation}
where the prime indicates differentiation with respect to the
independent variable $x$.  Equation (\ref{2.1}) arises in general
relativity. On the assumption of a conformal Killing vector of
a particular type for a spherically symmetric shear-free
gravitational field Dyer, McVittie and Oattes \cite{Dyer} obtained
the third order field equation
 \begin{equation}
 \mu^2T_{\mu\mu\mu} +\mu (2m - 1)T_{\mu\mu} + \left(m^2 - 2m + 2T\right)T_{\mu} = 0. \label{2.2}
 \end{equation}
Under the transformation
 \begin{equation}
 \mu = \exp\left(\frac{x}{2m - 4}\right),\qquad T (\mu) =
2 (m - 2)^2y (x) -\frac 12 \left(m^2-4m + 3\right) \label{2.3}
 \end{equation}
we obtain (\ref{2.1}) from (\ref{2.2}).  To determine the leading order
behaviour of (\ref{2.1}) we make the substitution $y = \alpha\chi^p
$, where $\chi = x-x_0 $ into (\ref{2.1}).  We find that the first
and the third terms are dominant, $\alpha = - 12 $ and $p = - 2$.  The substitution of
 \begin{equation}
 y = - 12\chi ^{- 2} +\beta\chi ^{r- 2} \label{2.4}
 \end{equation}
 into the two dominant terms gives the resonances $r = - 1,4,6 $.  We substitute
 \begin{equation}
 y = - 12\chi ^{- 2} +\sum_{i = - 1}^4 a_i\chi ^{i} \label{2.5}
 \end{equation}
into the full equation~(\ref{2.1}) to see it is
compatible at the resonances.  We find that
 \begin{equation}
a_{- 1} = \frac{12}{5},\qquad a_0 = \frac{1}{25}\qquad\mbox{\rm
and}\qquad a_1 = \frac{1}{125}  \label{2.6}
 \end{equation}
for the coefficients up to the first resonance at $r = 4 $.  At
the first resonance we have compatibility and, by construction,
$a_2 $ is arbitrary.  The next coefficient is given by
 \begin{equation}
 a_3 = \frac{1+ 1250a_2 }{18750}. \label{2.7}
 \end{equation}
 However, at the resonance, $r = 6 $, the condition of compatibility requires that
 \begin{equation}
 a_2 = -\frac{1}{1250}, \label{2.8}
 \end{equation}
i.e. the coefficient $a_3 $ is zero.
 Consequently (\ref{2.1}) fails the Painlev\'e test.  The solution requires
logarithmic terms to be introduced into the Laurent expansion and
this means that the general solution has infinitely many branches.
Infinite branching is strongly connected with nonintegrability and
this suggests that (\ref{2.1}) could exhibit
chaos~\cite{Chang}\cite[p.~348]{Tabor}\footnote{The logarithmic singularity of
the general solution can be avoided by accepting the constraint,
(\ref{2.8}), on the coefficient $a_2$ so that one obtains the series
expansion, $y = - 12\chi ^{- 2} +\frac{12}{5}\chi ^{- 1}+{25} +
\frac{1}{125}\chi +\frac{1}{1250}\chi^2+a_4\chi^4+\cdots$. This
expansion contains only the two free parameters, $x_0 $ and $a_4 $
and is interpreted to be the local representation of a particular
solution of (\ref{2.1}) in the sense of Cotsakis and Leach
\cite{Cotsakis,Flessas}.  There is another interpretation which
is that this series expansion is the general solution of a second
order equation which, in some sense, is an invariant of the third
order equation, (\ref{2.1}).}.  Analysis of the numerical properties of the
equation reveals that there is instability in the numerical solution but not
chaos.

Even a cursory examination of (\ref{2.1}) reveals the existence of a first integral
 \begin{equation}
 I (y,y',y'') = y''+y'+\frac 12 y^2 \label{2.10}
 \end{equation}
 which is associated with the obvious Lie point symmetry of (\ref{2.1}),
{\it videlicet} $\partial_x$. Instead of viewing (\ref{2.10}) as a first integral of (\ref{2.1})
one can consider it as an equation of the second order containing
a parameter, $I $, {\it videlicet}
 \begin{equation}
 y''+y'+\frac 12  y^2-I = 0. \label{2.11}
 \end{equation}
 This equation can be written as a family of two-dimensional first order
 systems,
\begin{gather}
\dot{x}   =y,\\
\dot{y}   =-y-\frac{1}{2}x^{2}+I.
\end{gather}
The equilibrium points of this system occur only for $I\geq0.$
They are $x=\pm\sqrt{2I},$ $y=0.$ A~simple application of the
Bendixon--Dulac criterion reveals that there are no limit cycles.
Thus, as $I$ goes from negative to positive values, the system
experiences a supercritical bifurcation with an additional
qualitative change in the nature of the critical points. The
equilibrium points on the positive $x$ axis are always saddle
points. On the other hand those lying on the negative $x$ axis
are stable nodes for $I<1/32$ and stable foci with clockwise
direction for $I>1/32.$

It is sometimes natural to consider the dynamics of a
two-dimensional family of systems depending upon parameters as
equivalent to that of a higher order model equation from which it
can derived. In our case, however, since the two-dimensional
reduced system has bifurcations and lines of equilibria it may
have completely different (and simpler) dynamics than that
hidden in the third order equation. In fact we can apply the
Painlev\'e test to (\ref{2.11}) and find that it satisfies the
necessary conditions for the possession of the Painlev\'e property
only in the case that $I = -18/625 $. Then it has the expansion
 \begin{equation}
y = - 12\chi ^{- 2} +\frac{12}{5}\chi ^{- 1}+{25} +
\frac{1}{125}\chi -\frac{1}{12500}\chi^2-\frac{1}{187500}\chi^3
+b_4\chi^4+\cdots, \label{2.12}
 \end{equation}
where $b_4 $ is the arbitrary constant introduced at the
resonance $r = 6 $.  The integrability of
 \begin{equation}
 y''+y'+\frac 12 y^2+\frac{18}{625} = 0 \label{2.13}
 \end{equation}
 has been demonstrated by Ince \cite{Ince} who transformed (\ref{2.13}) to
 \begin{equation}
 \frac{{\mbox{d}}^2w}{\mbox{d}z^2} = 6w^2, \label{2.14}
 \end{equation}
the solution of which can be expressed in terms of elliptic
functions, by means of the Kummer--Liouville transformation~\cite{Kummer, Liouville}
 \begin{equation}
y (x) = \exp\left(-\frac{2z}{5}\right)w (x) +\frac{6}{25},\qquad x = -
\frac{ 5i}{\sqrt{12}}\exp\left(-\frac{z}{5}\right). \label{2.15}
 \end{equation}
Equation (\ref{2.14}) is integrable in the sense of Painlev\'e and the transformation
(\ref{2.15})
is single-valued.  Hence (\ref{2.13}) is integrable.  Whatever
happens to the solutions of (\ref{2.1}) which commence at some point
not on the surface defined by (\ref{2.11}) in the phase space with
the particular value of $I = - 18/625 $ is not obvious, but those
which do are regular and remain on the surface.

As a final note on the properties of our model equation we remark
that the equation, (\ref{2.13}), has the additional Lie point
symmetry (calculated using Program {\tt Lie} \cite{Head1, Head2}),
$\exp\left(\frac{x}{5}\right)\left[\partial_x -\left(\frac 25 y-\frac{12}{125}
\right)\partial_y\right]$, in addition to the symmetry, $\partial_x$, possessed for
general values of the parameter, $I$.  The algebra of the
symmetries is Type~III of Lie's classification of two-dimensional
algebras \cite[p.~430, Kap.~19]{Lie} which has the normal form
 \begin{equation}
 G_1 = \partial_y,\qquad G_2 = x\partial_x +y\partial_y \label{2.16}
 \end{equation}
so that $[G_1,G_2] = G_1 $ and the normal form of the equation
invariant under the two symmetries (\ref{2.16}) is
 \begin{equation}
 XY''= F (Y'). \label{2.17}
 \end{equation}
 The tidy form of the equation, (\ref{2.14}), arises from using the nonnormal form
 \begin{equation}
 G_1 = \partial_x,\qquad G_2 = -x\partial_x +y\partial_y. \label{2.18}
 \end{equation}
The second symmetry in (\ref{2.16}) should not be expected to be
related to the complete symmetry group to be computed since (\ref{2.13})
 is not equivalent to (\ref{2.1}) as the particular value of $I $ brings
  us into the realm of configurational invariants~\cite{Hall,Sarlet}.

\section{Lie symmetries}
Equation (\ref{2.1}) has just the one Lie point symmetry, $\partial_x $,
which cannot be the complete symmetry group of (\ref{2.1}) since the
equation
 \begin{equation}
 y'''= f (y,y',y'') \label{3.2}
 \end{equation}
is the general form of a third order ordinary differential equation
invariant under $\partial_x$\footnote{One could consider introducing the idea of degrees of
completeness in symmetry groups.  This would depart from the
criterion of Krause that the equation be unique.  However, there
is no reason why we should not elaborate on Krause's proposition.}.
Examination of (\ref{2.1}) for contact symmetries shows that there
are none. Consequently the complete symmetry group (we now speak
strictly in the sense of Krause) must comprise nonlocal or
generalised symmetries.  To calculate these we follow the
procedure outlined in Pillay {\it et al}~\cite{Pillay} and used with
great effect by Bouquet {\it et al}~\cite{Bouquet} in their study of
second order ordinary differential equations possessing integrating factors~\cite{Roche}.

When one studies symmetries of differential equations without
placing any restrictions upon the coefficient functions one may
replace the differential operator
 \begin{equation}
 G = \xi\partial_x +\eta\partial_y \label{3.3}
 \end{equation}
 with the equivalent operator
 \begin{equation}
 \bar{G} = \bar{\eta}\partial_y,  \label{3.4}
 \end{equation}
where $\bar{\eta} =\eta -y'\xi $.  The advantage of using (\ref{3.4})
is that the calculations are simpler.  Dropping the overbar we
find that the action of the third extension of (\ref{3.4}) on
(\ref{2.1}) gives
 \begin{equation}
 \eta'''+\eta''+\eta'y+\eta y'= 0. \label{3.5}
 \end{equation}
Our task is to find a sufficient number of independent functions,
$\eta $, which are solutions of (\ref{3.5}) to be able to define
(\ref{2.1}) uniquely and so constitute the required number of
symmetries to make a complete symmetry group.

 It is a trivial matter to integrate (\ref{3.5}) once.  We have
 \begin{equation}
 \eta''+\eta'+\eta y = C, \label{3.6}
 \end{equation}
where $C $ is an arbitrary constant of integration.  We replace the
coefficient of $\eta $ from the original differential equation,
(\ref{2.1}), to obtain
 \begin{equation}
 (\eta''+\eta')y'- \eta (y'''+ y'') = Cy' \label{3.7}
 \end{equation}
 which can be integrated by parts to give
 \begin{equation}
 \eta'y'-\eta y''+\int ( \eta'y'-\eta y'')\mbox{d} x = Cy +K, \label{3.8}
 \end{equation}
where $K $ is another arbitrary constant.  If we let $\omega =
\int( \eta'y'-\eta y'')\mbox{d} x $, the integro-differential equation
(\ref{3.8}) becomes the first order linear nonhomogeneous equation
 \begin{equation}
 \omega'+\omega = Cy +K \label{3.9}
 \end{equation}
 which has the solution
 \begin{equation}
 \omega = J{\mbox{e}} ^{-x} +{\mbox{e}}^{-x}\int (Cy +K){\mbox{e}}^x\mbox{d} x, \label{3.10}
 \end{equation}
 where $J $ is a constant of integration, so that $\eta $ now satisfies its
own linear first order nonhomogeneous equation,
 \begin{equation}
 \eta'y'-\eta y''= Cy +K-J{\mbox{e}} ^{-x} -{\mbox{e}}^{-x}\int (Cy +K){\mbox{e}}^x\mbox{d} x. \label{3.11}
 \end{equation}
 The solution of (\ref{3.11}) is
 \begin{equation}
\eta = Ay'+y'\int\frac{1}{y'{}^2}\left[ Cy +K- J{\mbox{e}} ^{-x} -{\mbox{e}}^{-x}
\int (Cy +K) {\mbox{e}}^x\mbox{d} x\right]\mbox{d} x, \label{3.12}
 \end{equation}
 where $A $ is a constant of integration.

In (\ref{3.12}) there are four constants, but one of them, $K $, is
spurious.  We have the three symmetries
\begin{gather}
 G_1 = y'\partial_y \ \Leftrightarrow \ G_1 = \partial_x, \nonumber\\
G_2 = y'\left(\int\frac{\mbox{d} x}{y'{}^2{\mbox{e}}^x }\right)\partial_y \ \Leftrightarrow \
G_2 = \left(\int\frac{\mbox{d} x}{y'{}^2{\mbox{e}}^x }\right)\partial_x, \nonumber\\
 G_3 = y'\left(\int\frac{\mbox{d} x}{y'{}^2}\left(y-{\mbox{e}} ^{-x}\int y{\mbox{e}}^x
\mbox{d} x\right)\right)\partial_y\nonumber\\
\qquad \ \Leftrightarrow \  G_3 = \left(\int\frac{\mbox{d} x}{y'{}^2}\left(y-{\mbox{e}} ^{-x}\int y{\mbox{e}}^x
\mbox{d} x\right)\right)\partial_x,\label{3.13}
 \end{gather}
where we have added the corresponding operator when we use only
$\partial_x $ for the symmetry.  We observe that the first of (\ref{3.13})
is just the obvious symmetry of (\ref{2.1}).

Our immediate task is to find the form of the third order
differential equation which is invariant under the actions of the
third extensions of these three symmetries.  We commence with the
general equation
 \begin{equation}
 y'''= f (x, y,y',y''). \label{3.14}
 \end{equation}
For the calculations we use the $\partial_x $ form of the symmetries.
Invariance under $G_1 $ implies that the equation have the form
 \begin{equation}
 y'''= f (y,y',y''). \label{3.15}
 \end{equation}
Invariance under $ G_2 $ and $G_3 $ is not obtainable by
inspection!  The third extension of $G_2 $ is
 \begin{equation}
G_2^{[3]} = \left(\int\frac{\mbox{d} x}{y'{}^2{\mbox{e}}^x }\right)\partial_x
-\left(\frac{1}{y'{}{\mbox{e}}^x }\right)\partial_{y'} +
\left(\frac{1}{y'{\mbox{e}}^x}\right)\partial_{y''} -\left[
\frac{y'''+y''+y'}{y'{}^2{\mbox{e}}^x}\right]\partial_{y'''} \label{3.16}
 \end{equation}
and, when it acts on (\ref{3.15}), we obtain the linear first order
partial differential equation
 \begin{equation}
 y'\frac{\partial f}{\partial y''} -y'\frac{\partial f}{\partial y'} = -f-y''-y' \label{3.17}
 \end{equation}
 which has the two characteristics
 \begin{equation}
 u = y''+ y'\qquad\mbox{\rm and}\qquad v = \frac{f+y'+y''}{y'} \label{3.18}
 \end{equation}
 so that the right side of (\ref{3.15}) has the form
 \begin{equation}
 f = -(y''+y') +y'g\left(y''+ y',y\right), \label{3.19}
 \end{equation}
 where $g $ is an arbitrary function of its arguments.

 The third extension of $G_3 $ is
 \begin{gather}
 G_3 ^{[3]} = \left(\int\frac{\mbox{d} x}{y'{}^2}\left(y-{\mbox{e}} ^{-x}\int
y{\mbox{e}}^x\mbox{d} x\right)\right)\partial_x
- \frac{1}{y'{} }\left(y-{\mbox{e}} ^{-x}\int y{\mbox{e}}^x\mbox{d} x\right)\partial_{y'}\nonumber\\
\phantom{ G_3 ^{[3]} =}  - \frac{1}{y'{} }\left(y'-y+{\mbox{e}}^{-x}\int y{\mbox{e}}^x\mbox{d}
 x\right)\partial_{y''} +
\partial_{y'''}. \label{3.20}
 \end{gather}

From the action of (\ref{3.20}) and (\ref{3.15}) with $f $ as given in
(\ref{3.19}) we find that $g $ is independent of its first argument
and so the general form of the third order equation invariant
under the three symmetries (\ref{3.13}) is
 \begin{equation}
 y'''+y''+y'+y'h (y) = 0, \label{3.21}
 \end{equation}
 where $h (y) $ is an arbitrary function.

It is quite evident from (\ref{3.21}) that the three symmetries
(\ref{3.13}), which arise from the direct integration of the linear third order
ordinary differential equation (\ref{3.5}), are insufficient to specify completely our model
equation, (\ref{2.1}), and so we must find some other symmetry. We
assume that the symmetry has the form $\xi\partial_x $.  The action of
the third extension of the symmetry on (\ref{2.1}) gives
 \begin{equation}
 y'\xi'''+\left(3y''+y'\right)\xi''+\left(2y'''+y''\right)\xi' = 0, \label{3.22}
 \end{equation}
in which (\ref{2.1}) has been used.  Equation (\ref{3.22}) is exact and
can be integrated once to give
 \begin{equation}
 y'\xi''+\left(2y''+y'\right)\xi'= A, \label{3.23}
 \end{equation}
where $A $ is a constant of integration, which is a linear first
order nonhomogeneous equation for $\xi'$ with the solution
 \begin{equation}
 \xi'= A\frac{\int y'{\mbox{e}}^x\mbox{d} x}{y'{}^2{\mbox{e}}^x} +
B \frac{1}{y'{}^2{\mbox{e}}^x} \label{3.24}
 \end{equation}
 and so
 \begin{equation}
 \xi = A \int\left(\frac{\int y'{\mbox{e}}^x\mbox{d} x}{y'{}^2{\mbox{e}}^x}\right)\mbox{d} x +
B \int\frac{\mbox{d} x}{y'{}^2{\mbox{e}}^x} +C,\label{3.25}
 \end{equation}
where $B $ and $C $ are further constants of integration.
Unfortunately, and this should not be surprising, the three symmetries
 in (\ref{3.25}) are just the three
symmetries listed in (\ref{3.13}).

Evidently we are not going to be able to obtain the necessary
additional symmetries the easy way and so we must assume that the
symmetry has the form ({\it cf} the {\it Ansatz} of Krause \cite{Krause})
 \begin{equation}
 G = \left(\int\xi\mbox{d} x\right)\partial_x +\eta\partial_y. \label{3.26}
 \end{equation}
 The action of the third extension of (\ref{3.26}) on the equation (\ref{2.1}) gives
 \begin{equation}
 y'\xi''+ y''\xi'+ (2y''+y')\xi'+ (2y'''+y'')\xi =\eta'''+\eta''+\eta'y+\eta y' \label{3.27}
 \end{equation}
 which is easily integrated to give
 \begin{equation}
\xi= \frac{1}{y'{}^2{\mbox{e}}^x} \left[ A\int y'{\mbox{e}}^x\mbox{d} x +B+\int
y'{\mbox{e}}^x\left(\eta''+\eta'+\eta y\right)\mbox{d} x\right]. \label{3.28}
 \end{equation}

In addition to the three symmetries of (\ref{3.25}) we have the
general symmetry
 \begin{equation}
 G = \int\left\{ \frac{1}{y'{}^2{\mbox{e}}^x} \left[\int y'{\mbox{e}}^x
\left(\eta''+\eta'+\eta y\right)\mbox{d}
x\right]\right\}\partial_x +\eta\partial_y \label{3.29}
 \end{equation}
for arbitrary functions $\eta $.  If we apply the third extension
of (\ref{3.29}) to the equation (\ref{3.21}), we obtain, after some
simplification,
 \begin{equation}
 \eta y'h'+\eta'h = (\eta y)'-\eta', \label{3.30}
 \end{equation}
in which the prime on $\eta $ represents differentiation with
respect to $x $ and the prime on $h (y) $ differentiation with
respect to $y $.  If we make use of the chain rule, we may write
(\ref{3.30}) as
 \begin{equation}
 (\eta h)' = (\eta y)'-\eta', \label{3.31}
 \end{equation}
 where now all primes represent differentiation with respect to $y $ and it is
evident that $\eta $ must be a function of $y $ only.  We integrate (\ref{3.31}) to obtain
 \begin{equation}
 h (y) = \frac{A}{\eta} + y- 1 \label{3.32}
 \end{equation}
 and so, up to a constant, we obtain (\ref{2.1}) if we take $\eta $ to be a constant.

 \section{Reconstruction of the model equation}

 The third order equation
 \begin{equation}
 y'''+y''+ Ky'+yy'= 0 \label{4.1}
 \end{equation}
 has the four symmetries
 \begin{gather}
 G_1 = \partial_x, \qquad
 G_2 = \left(\int\frac{\mbox{d} x}{y'{}^2{\mbox{e}}^x }\right)\partial_x,\nonumber\\
 G_3 = \left(\int\frac{\mbox{d} x}{y'{}^2}\left(y-{\mbox{e}} ^{-x}\int
y{\mbox{e}}^x\mbox{d} x\right)\right)\partial_x,\nonumber\\
 G_4 = \int\left\{ \frac{1}{y'{}^2{\mbox{e}}^x} \left[\int yy'
{\mbox{e}}^x\mbox{d} x\right]\right\}\partial_x +\partial_y \label{4.2}
 \end{gather}
which specify it completely.  This, of course, is not the model
equation with which we started and we would do well to verify
that the presence of the additional term makes no difference to
the behaviour of the solution of the equation.

When we perform the Painlev\'e analysis we find that the first
and fourth terms in (\ref{4.1}) are dominant, the singularity is a
double pole and the resonances are $r = - 1,4 $ and $6 $ as for
the model equation, (\ref{2.1}).  When we make the substitution
 \begin{equation}
 y = \sum_{i  = 0}^6a_i\chi ^{i- 2} \label{4.3}
 \end{equation}
 to check for compatibility, we find that
 \begin{equation}
a_0 =-12,\qquad a_1 = \frac{12}{5},\qquad a_2 = \frac{6}{25}
-K,\qquad a_3 = \frac{6}{125}, \label{4.4}
 \end{equation}
but then, at the resonance $r = 4$,  $a_4 = 0 $ and so there can be
no Painlev\'e property for (\ref{4.1}).

 Equation (\ref{4.1}) is trivially integrated to give
 \begin{equation}
 y''+y'+ Ky+\frac 12 y^2-I = 0, \label{4.5}
 \end{equation}
where  $I $ is the value of the first integral.  When we perform
the Painlev\'e analysis on (\ref{4.5}), we find that the singularity
is a double pole and the resonances are at $r = - 1 $ and $6 $.
We make the same substitution as in (\ref{4.3}) and find that the
coefficients are given by
 \begin{gather}
 a_0 = - 12,\qquad a_1 = \frac{12}{5},\qquad a_2 = \frac{1}{25} -K,
\qquad a_3 = \frac{1}{125},\nonumber\\
a_4 = \frac{7}{2500} -\frac{2I + K^2}{20},\qquad
a_5 = \frac{158}{25 000} -\frac{11 (2I + (K+ 1)^2)}{300},
 \label{4.6}
 \end{gather}
 that $a_6 $ is arbitrary and that there is compatibility at this resonance provided
 \begin{equation}
 2I + K^2 = -\frac{36}{625}. \label{4.7}
 \end{equation}

Finally we compute the Lie point symmetries of (\ref{4.5}) when the
constants are related according to (\ref{4.7}).  They are
 \begin{equation}
G_1 = \partial_x\qquad\mbox{\rm and}\qquad G_2 = {\mbox{e}} ^{x/5}\left[\partial_x
+\frac{2}{5}\left(\frac{6}{25} -K-y\right)\partial_y\right].
 \label{4.8}
 \end{equation}
We observe that these features are identical to those of (\ref{2.1})
when we make the substitution $K = 0 $.  It is not feasible to
make numerical experiments for arbitrary $K $, but we are
confident that the behaviour of the solutions of (\ref{4.1}) is the
same as the solutions of our original model equation.

 We can demonstrate a basis for that confidence by using the symmetries to
reduce the order of (\ref{4.1}).  Naturally we commence with the only point
symmetry, $G_1 $, which has the invariants $u = y $ and $v = y'$.  Equation
(\ref{4.5}) becomes
 \begin{equation}
 vv''+v'{}^2+v'+K+u = 0 \label{4.9}
 \end{equation}
 and the other symmetries become
 \begin{gather}
G_2 \rightarrow \left(\frac{1}{v}\exp\left[ -\int\frac{\mbox{d} u}{v}\right]\right)\partial_v, \nonumber\\
G_3 \rightarrow \left(\frac{1}{v}\exp\left[ -\int\frac{\mbox{d} u}{v}\right]\int\exp\left[
\int\frac{\mbox{d} u}{v}\mbox{d} u\right]\right)\partial_v, \nonumber\\
G_4 \rightarrow \partial_u -\left(\frac{1}{v}\exp\left[ -\int\frac{\mbox{d} u}{v}\right]
\int u\exp\left[\int\frac{\mbox{d} u}{v}\mbox{d} u\right]\right)\partial_v.
 \label{4.10}
 \end{gather}
We observe that $G_2 $ has become an exponential nonlocal
symmetry and consequently can be used to reduce the order of
(\ref{4.9})~\cite{Feix}. We obtain the invariants $z = u $ and $w =
vv'+v $ and (\ref{4.9}) is reduced to the first order equation
 \begin{equation}
 \frac{\mbox{d} w}{\mbox{d}z} = - (K+z). \label{4.11}
 \end{equation}
We can now see that the presence of $K $ has no bearing upon the
behaviour of the solution of the equation except as a translation
of the initial condition on $y (x_0) $, thereby justifying our
comment above.  Under this reduction of order the two unused
symmetries become
 \begin{equation}
G_3 \rightarrow \partial_w, \qquad
G_4 \rightarrow z\partial_w \label{4.12}
 \end{equation}
 and so we can infer that these two symmetries have the Abelian algebra $2A_1 $.

As (\ref{4.11}) has two Lie point symmetries, it is obviously
integrable.  We obtain
 \begin{equation}
 w (z) = C-\frac 12 (K+z)^2 \label{4.30}
 \end{equation}
 so that the next equation to be solved is
 \begin{equation}
 vv'+v = C-\frac 12 (K+u)^2. \label{4.13}
 \end{equation}
Equation (\ref{4.13}) is an Abel's equation of the second kind
which, not surprisingly, does not yield to any of the methods of
solution proposed by Kamke \cite[p.~24, Art 4.10 and 4.11]{Kamke}.

 In the integrable case the second order equation to be solved is
 \begin{equation}
 y''+y'+\frac 12 (y+K)^2-\frac{18}{625} = 0. \label{4.14}
 \end{equation}
 This can be transformed to an equation of Emden--Fowler type
 \begin{equation}
 \ddot{w} +w^2 = 0 \label{4.15}
 \end{equation}
  by means of the Kummer--Liouville transformation
 \begin{equation}
 y = 2{\mbox{e}}^{2x/5}w (t) -K-\frac{6}{25},\qquad t'(x) = {\mbox{e}}^{-x/5}. \label{4.16}
 \end{equation}
The solution of (\ref{4.15}) is in terms of elliptic functions and so
the solution of (\ref{4.14}) is in terms of elliptic functions of
somewhat complicated argument.  It is interesting that on the
surface in the space of initial conditions specified by the value
of the integral this solution is analytic except for the polelike
singularity and yet just off this surface one may find
nonintegrable behaviour.

We can always make use of a Kummer--Liouville transformation to
bring (\ref{4.5}) into the form of a generalised Emden--Fowler
equation.  If we put
 \begin{equation}
 y+K =u(x)v(t) +\alpha\qquad\mbox{\rm and}\qquad t = t (x), \label{4.17}
 \end{equation}
(\ref{4.5}) becomes
 \begin{gather}
\left(ut'{}^2\right)\ddot{v} +\left(2u't'+ut''+ut'\right)\dot{v}\nonumber\\
\qquad {}+\left(u''+u'+u\alpha\right)v+ \frac 12 u^2v^2+\frac 12 \alpha^2 - \frac 12 \beta^2 = 0,
 \label{4.18}
 \end{gather}
where we have written $2I +K^2 = \beta^2 $.  We remove the
constant term by setting $\alpha =\beta $.  We remove the
coefficient of $v $ by solving the equation
 \begin{equation}
 u''+u'+u\beta = 0. \label{4.19}
 \end{equation}
The coefficient of $\dot{v} $ is removed if we set $t'= u ^{-
2}{\mbox{e}}^{-x} $ and we are left with the equation
 \begin{equation}
 \ddot{v} +\phi (t)v^2 = 0, \label{4.20}
 \end{equation}
 where
 \begin{equation}
 \phi (t (x)) = \frac 12{\mbox{e}}^{-x/2}\left(A{\mbox{e}}^{\gamma x} +
B{\mbox{e}}^{-\gamma x}\right)^5, \label{4.21}
 \end{equation}
$A $ and $B $ are constants of integration and $\gamma =
\sqrt{\frac 14 - (2I +K^2) ^{1/2}} $.  We note that the form of $\phi
(t (x)) $ is not one of the functions for which the generalised
Emden--Fowler equation~(\ref{4.20}), has been found to be integrable
\cite{Leach3, Mellin}.

\section{Discussion}

As in the case of Krause's analysis for the complete symmetry
group of the Kepler problem, we can specify our model equation up
to a constant.  The Kepler problem, being a three-dimensional
system of second order equations, required eight symmetries, three
of which were nonlocal.  Our scalar third order equation is
similarly specified with four symmetries, three of which are
nonlocal.

 Already in the paper by Krause \cite{Krause} there was some concern as to the
proper way to consider the algebraic properties of the symmetries, particularly
the nonlocal symmetries.  In fact he comments ``these are not point
transformations and one does not obtain from them a set of differential
operators satisfying a Lie algebra, as in the standard manner.'' \cite[p. 5738]{Krause}.
The method used by Krause to overcome this obstacle was to look at
the transformations generated by the symmetries in the coordinates and the
rate of change of local time.  In both of these the nonlocality of the
symmetries did not appear and he was able to construct a satisfactory group
representation.

 Our approach in this matter is somewhat different because we attempt
to obtain a representation of an algebra from the symmetries under
consideration and consequently we do not want to have the ambiguities of the
algebras of nonlocal symmetries to intrude as an unnecessary complicating
factor.  Our aim is to present the concept in the simplest possible context.
We also consider the question of the
uniqueness of the complete symmetry group, because we do not believe that it
is unique.  Rather one would expect there to be equivalent representations and
we demonstrate that with the great paradigm, the
 one-dimensional free particle.  The free particle in one dimension with the equation
 \begin{equation}
 \ddot{x} = 0 \label{4.22}
 \end{equation}
 has the eight Lie point symmetries
\begin{align}
& G_1 = \partial_x , && G_2 = t, \partial_x, && G_3 = x\partial_x, &&
G_4 = \partial_t,\nonumber\\
& G_5 = 2t\partial_t +x\partial_x, && G_6 = t^2\partial_t + tx\partial_x,
 && G_7 = x\partial_t, &&G_8 = tx\partial_t + x^2\partial_x \label{4.23}
 \end{align}
with the algebra $sl (3,R) $ which has the substructure $sl
(2,R)\oplus_s\left\{2 A_1\oplus A_2\oplus A_1\right\} $.  If we
assume the general form of a second order equation,
 \begin{equation}
 \ddot{x} = f (t,x,\dot{x}), \label{4.24}
 \end{equation}
 we find, for example, that $f $ must take the value in (\ref{4.22}) with the
combinations of symmetries $G_1 $, $G_2 $ and $G_3 $; $G_3$, $G_7$ and $G_8$.
  The algebra of the each triple is
$A_{3,3} $ ($\Leftrightarrow D\oplus_s T_2 $, the semidirect sum of dilations
and translations).  The triple  $G_1 $, $G_7 $ and $G_8 $
also completely specifies the equation, but it does not close.
To obtain a closed algebra one must add the symmetries $G_3 $,
$G_4 $ and $G_5 $.  The triple $G_4 $, $G_5 $ and
$G_6 $, constituting the subalgebra $sl(2,R)$ and characteristic of all
differential equations of maximal symmetry, gives the Ermakov--Pinney equation
\begin{equation}
\ddot{x} = \frac{K}{x^3} \label{4.25}
\end{equation}
and needs an additional symmetry, say $G_1$, to recover (\ref{4.22}).

This example demonstrates quite clearly the possible
nonuniqueness of the complete symmetry group since these three algebras are
quite distinct and cannot be different realisations of the same group because
they are of different dimension.

We conclude that we have demonstrated that even an ordinary
differential equation which  has Painlev\'e properties consistent
with nonintegrable behaviour is defined in terms of its
symmetries. We do not believe that our model equation, (\ref{2.1}),
is exceptional and propose that all ordinary differential
equations, be they derived in a Newtonian or relativistic context or elsewhere,
can be defined in
terms of a set of Lie symmetries. We repeat the point made earlier
at the conclusion of Section~2. The Lie symmetries of a
differential equation both provide a means to identify the
specific structure of the equation and a possible route towards
its solution.  The second property is not automatically available
in any given symmetry. For a symmetry to be part of the route to
the solution of a differential equation the symmetry must be able
to provide a means for the reduction of order, if not immediately,
at some level in the reduction process.

 \subsection*{Acknowledgements}

  MCN thanks the MURST (Cofin 97: Metodi e applicazioni di equazioni
differenziali ordinarie) for its support and PGLL expresses his
deep appreciation of the hospitality of the Dipartimento di
Matematica e Informatica, Universit\`a di Perugia, during the
period in which this work was initiated, thanks the Dean of the
School of Sciences, University of the Aegean, Professor G P
Flessas, and the Director of GEODYSYC, Dr S Cotsakis, for their
kind hospitality while this work was undertaken and the National
Research Foundation of South Africa and the University of Natal
for their continuing support.

\label{leach-lastpage}
\end{document}